\documentclass[aip,twocolumn,reprint]{revtex4-1}

\draft 

\usepackage{wrapfig}
\usepackage[]{graphicx,xcolor}
\usepackage{tabularx}
\usepackage{booktabs}
\usepackage{textcomp}
\usepackage{amsmath}
\usepackage{amssymb}
\usepackage[]{graphics}
\usepackage{amssymb}
\usepackage{amsmath}
\usepackage{color}
\usepackage{ifpdf}
\usepackage{lipsum}
\usepackage{siunitx}
\usepackage{braket}

\graphicspath{{figs/}} 
\newcommand{\executeiffilenewer}[3]{%
\ifnum\pdfstrcmp{\pdffilemoddate{#1}}%
{\pdffilemoddate{#2}} > 0 {\immediate\write18{#3}}\fi}

\ifpdf\usepackage{epstopdf}\fi

\begin{document}


\title{Magnetic field-free circularly polarized thermal emission from chiral metasurface} 



\author{S.~A.~Dyakov}
\email[]{e-mail: s.dyakov@skoltech.ru}
\affiliation{Skolkovo Institute of Science and Technology, 143025 Moscow Region, Russia}

\author{V.~A.~Semenenko}
\affiliation{Skolkovo Institute of Science and Technology, 143025 Moscow Region, Russia}

\author{N.~A.~Gippius}
\affiliation{Skolkovo Institute of Science and Technology, 143025 Moscow Region, Russia}

\author{S.~G.~Tikhodeev}
\affiliation{A.~M.~Prokhorov General Physics Institute, RAS, Vavilova 38, Moscow, Russia}
\affiliation{Faculty of Physics, Lomonosov Moscow State University, 119991 Moscow, Russia}


\date{\today}

\begin{abstract}
Thermal radiation from bulk disorderly placed nonresonant emitters is incoherent, broadband and isotropic. In an external magnetic field the thermal radiation from any source is circularly polarized. Here we propose a thermal radiation source which emits circularly polarized radiation and which is not placed in a magnetic field. The thermal source consists of a slab waveguide with etched chiral metasurface. Due to the absence of a mirror symmetry of the metasurface, the thermally generated electromagnetic waves become circularly polarized. In this letter we discuss the origin of this phenomenon in details. Using the Fourier modal method we analyze the eigenmodes of the structure and the emissivity spectra. We demonstrate that the degree of circular polarization in an optimized structure can be as high as 0.87.
\end{abstract}


\maketitle 

\section{Introduction}
In the last years, the study of far field and near field thermal emission of arteficial materials  attracted a great deal of attention from researchers due to its high potential for important applications in near-field thermal management \cite{ben2017thermal, kim2015radiative, ghanekar2018near, ben2016towards, ben2016photon, sa2015near, song2016radiative, dyakov2015, ben2015heat, PhysRevB.90.045414, singer2015near, PhysRevB.92.035419, lin2017near, mirmoosa2016microgap, lim2018optimization, yang2017electrically, basu2015near, PhysRevB.94.125431, guo2013thermal, PhysRevB.93.155403}, energy harvesting, and coherent thermal sources \cite{greffet2002coherent, guo2012broadband, wang2013measurement, pipa2013thermal, maruyama2001thermal, marquier2004coherent, sai2003high, kruk2016magnetic, narayanaswamy2004thermal, biswas2006theory, kats2013vanadium, sai2005tuning, sai2001spectral, celanovic2005resonant}. Photonic crystal surfaces were demonstrated to be an effective tool to control the spectral, angular and coherence characteristics of thermal radiation \cite{greffet2002coherent, sai2003high, sai2001spectral, Lee:08, ueba2012spectral}. The angular emission diagram and the polarization of thermal radiation are dictated by the emitter symmetry. In particular, a structure which lacks a mirror symmetry can generate circularly polarized thermal radiation. The mirror symmetry can be broken down by applying an external magnetic field due to spin-orbit interaction of electrons \cite{argyres1955theory}. This phenomenon is known as magneto-optic Kerr effect and explains the strong circular polarization of white dwarfs emission \cite{kemp1970discovery}. The circularly polarized thermal radiation in a magnetic field has been observed in a laboratory too [see for example Refs.\,\onlinecite{kemp1970circular,kollyukh2005magnetic}]. Another way to break down the mirror symmetry is creating a structure with a chiral morphology. Chiral metasurfaces were used in Refs.\,\onlinecite{konishi2011circularly, lobanov2015polarization, lobanov2015controlling, demenev2016circularly, PhysRevB.89.045316} to obtain the circularly polarized photoluminescence of semiconductor quantum dots. The degree of circular polarization (DCP) depends on the surface geometry and is a matter of theoretical optimization. The highest obtaned degree of circular polarization was close to 100\% in the optimized structures.
\begin{figure}[b!]
\centering
\includegraphics[width=0.8\columnwidth]{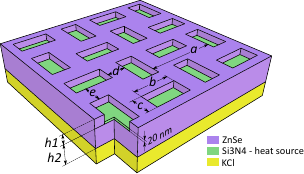}
\caption{(Color online) Schematic of chiral metasurface. The period of the photonic crystal structures is $a=10$\,$\mu$m, $b=2a/5$, $c=a/5$.}
\label{sample}
\end{figure}

In this paper we use the concept of a chiral metasurface to generate the circularly polarized thermal radiation. We propose a structure with artificial shape-induced chirality which radiates the circularly polarized thermal emission and discuss this effect in details.


\begin{figure*}[thp]
\centering
\includegraphics[width=1\textwidth]{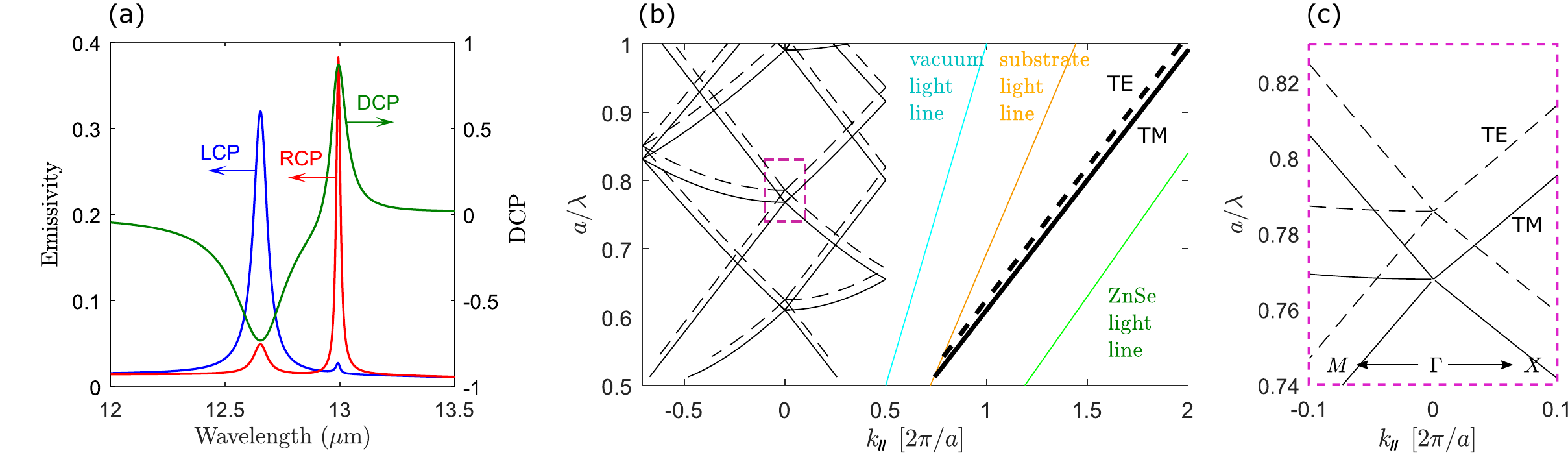}
\caption{(Color online) (a) LCP (blue) and RCP (red) emissivity spectra of the chiral metasurface with $h_1=4$\,$\mu$m and $h_2=5$\,$\mu$m. Degree of circular polarization is shown by green. (b): The lowest TE and TM waveguided modes in an effective homogeneous uniaxial double slab waveguide and the same modes folded into the first Brillouin zone. The dispersion of folded modes is shown along $\Gamma$-X and $\Gamma$-M directions. $h_1=4$\,$\mu$m, $h_2=5$\,$\mu$m. The modes bounded by the magenta dashed rectangle are shown in panel (c). Arrows show the corresponding photonic crystal directions.}
\label{fig2}
\end{figure*}

\section{Model structure and theory}
We propose a thermal emitter consistsing of a KCl substrate capped by ZnSe waveguide with two-dimensional array of etched rectangles (Fig.\,\ref{sample}). The etched pattern has chiral morphology with $C_4$ rotational symmetry. The bottom surface of rectangles is covered by 20-nm thick layer of Si$_3$N$_4$. In this work we assume that the temperature of the thermal emitter is close to 300\,K and hence we are focused on the 7--15\,$\mu$m wavelength range. The choice of materials ZnSe and KCl is attributed to the fact that they are transparent in the middle infrared and hence do not contribute to the thermal emission in this spectral range. In contrast, Si$_3$N$_4$ has a wide absorption band and, therefore, is the only source of thermal radiation in the structure. Dielectric permittivities of all of the above materials have weak dispersion and therefore in calculation we consider them dispersionless in the wavelength range of interest: $\varepsilon$(ZnSe)=5.67, $\varepsilon$(KCl)=2.08 and $\varepsilon$(Si$_3$N$_4$)=10.5+9.2$i$.

In this paper, the emissivity is calculated by the Kirhoff's law which states that the absorptivity and emissivity are equal. This has been numerically verified for uniform and photonic crystal slabs\cite{luo2004thermal}. In turn, the absorptivity is calculated using the Fourier modal method in the scattering matrix form \cite{Tikhodeev2002b, moharam1995formulation, whittaker99, messina2017radiative}. The decompositions of electric and magnetic fields into Fourier series  were done using Li's factorization rules~\cite{li1996use} with $13\times 13 = 169$ spatial harmonics. We checked the accuracy of the method with up to $31\times 31 = 961$ spatial harmonics for most important results.

\section{Results and discussions}

\begin{figure}[thp]
\centering
\includegraphics[width=0.75\linewidth]{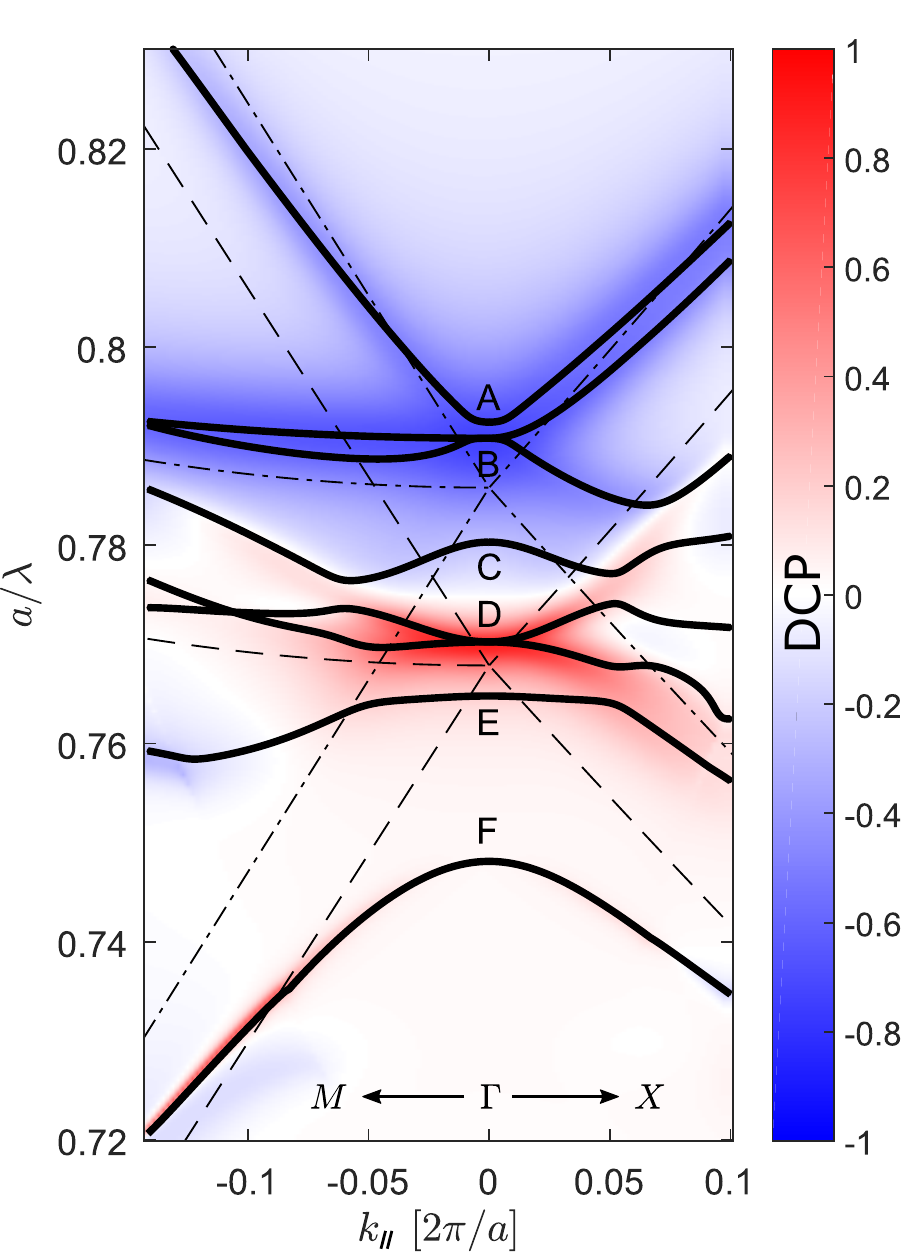}
\caption{(Color online) The calculated dispersion of quasiguided modes octet along  $\Gamma-M$ and $\Gamma-X$ directions near $\Gamma$ point (solid lines) against the background of the energy and $k_x$ dependence of the DCP. Color scale is explained on the right. Dashed and dash-dotted lines represent the structure TM and TE eigenmodes in empty lattice approximation. Arrows show the corresponding photonic crystal directions.}
\label{disp}
\end{figure}

\begin{figure*}[t!]
\centering
\includegraphics[width=1\textwidth]{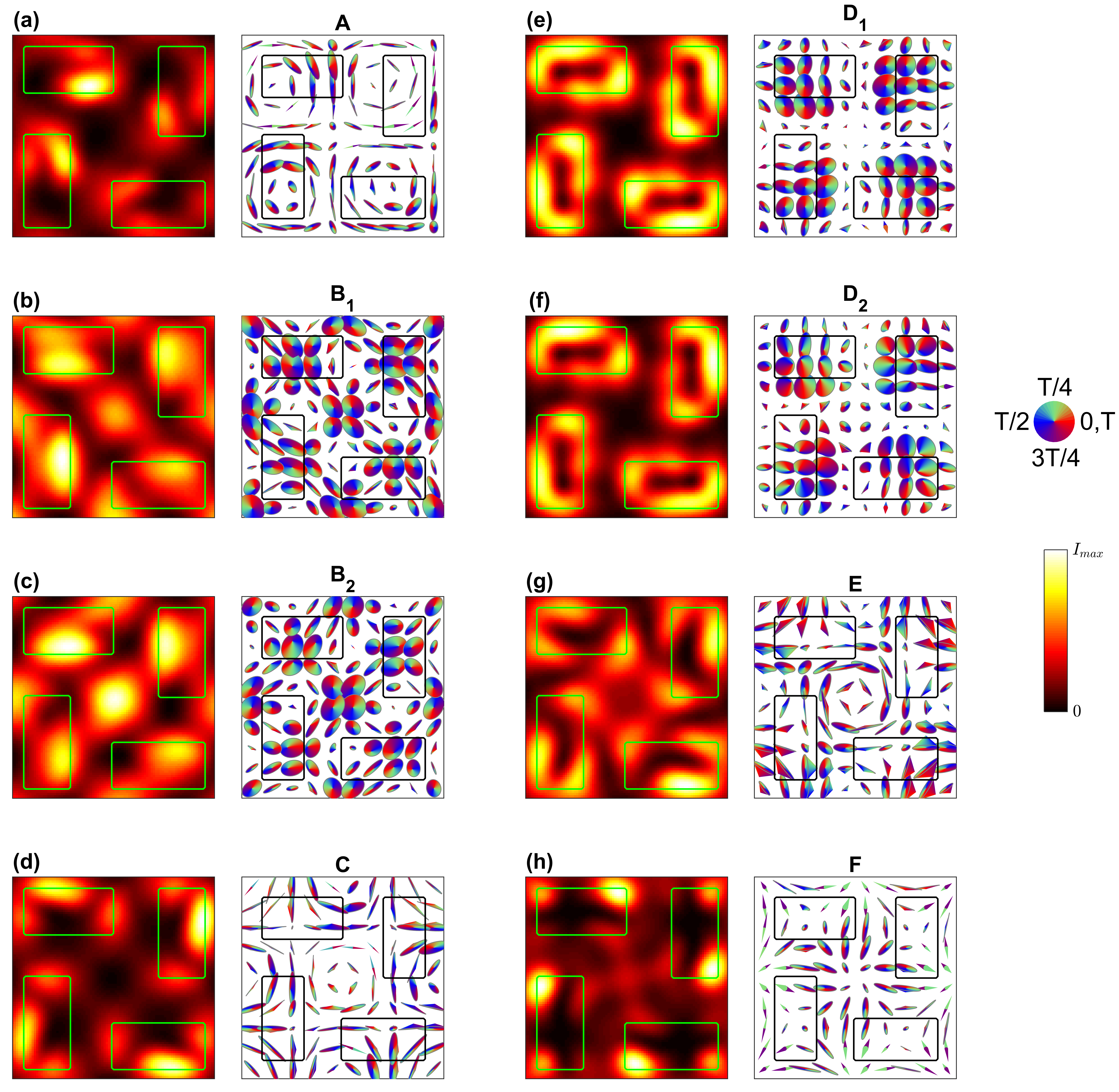}
\caption{(Color online) Left panels: electric field intensities of the quasiguided modes A--F in $\Gamma$ point. The maximal field intensities over the displayed cross sections normalized to those in free space are shown in titles. Right panels: {phase representation} of the fields showing their polarization characteristics. Fields are calculated in the middle of the absorbing layer. Green and black lines are material boundaries.}
\label{efields}
\end{figure*}

To estimate the expected circular polarization of thermal emission, we calculate the emissivity spectra of chiral metasurface in left circular polarization (LCP) $I_\mathrm{LCP}$ and right circular polarization (RCP) $I_\mathrm{RCP}$ as well as the degree of circular polarization. We estimate the DCP as 
\begin{equation}
\rho_c = \frac{I_\mathrm{RCP}-I_\mathrm{LCP}}{I_\mathrm{RCP}+I_\mathrm{LCP}}
\end{equation}
As shown in Fig.\,\ref{fig2}a, in the displayed spectral range, the emissivity is characterized by the two peaks both having different amplitudes in LCP and RCP. As a result, the DCP is non-zero and reaches the values of -0.73 and 0.87 at $\lambda=12.65$ and 13\,$\mu$m correspondingly. 

To understand the physical origin of the emissivity peaks, we analyse the structure eigenmodes. At first, we use the simplest approach for this, namely, an empty lattice approximation. In this approximation we replace the photonic crystal layer by a uniaxial homogeneous medium (UHM) with diagonal dielectric permittivity tensor. For the sake of simplicity we omit the thin Si$_3$N$_4$ layer and replace the rectangular holes in ZnSe layer by the cylindric ones. In this simplified approximation we lose the chirality of the structure (the symmetry becomes $C_{4v}$ instead of $C_{4}$) and thus cannot estimate the DCP. But it allows us at least to approximate the energy dispersion of the TE and TM waveguide modes in the structure. For this purpose we roughly estimate the effective ordinary and extraordinary dielectric permittivities of UHM from the generalized Bruggeman formula:
\begin{equation}
\sum_{j=1}^2f_j\frac{\varepsilon_\mathrm{eff}-\varepsilon_{j}}{\varepsilon_\mathrm{eff}+L(\varepsilon_{j}-\varepsilon_\mathrm{eff})}=0,
\label{emt}
\end{equation}
where index $j=1,2$ denotes air or ZnSe, $\varepsilon_j$ is the dielectric permittivity, $f_j$ is the filling factor, and $L$ the depolarization factor. We consider the entire effective structure as a dielectric double slab waveguide and find its guided modes. At $k_z=0$ in the TE polarized guided modes, the electric field is perpendicular to the z-axis, while in TM polarization the electric field is approximately parallel to the z-axis. Thus, in equation ({\ref{emt}}) we use the depolarization factors typical for the cylindrical inclusions with cyliner axis oriented along the z-axis: $L_\mathrm{TE}=1/2$ and $L_\mathrm{TM}=0$. The effective dielectric permittivities, obtained from Eq.\,\ref{emt}, are \footnote{We notice, however, that $\varepsilon_\mathrm{eff, TM} = 4.2$ describes the optical resonances of the chiral metasurface better than $\varepsilon_\mathrm{eff, TM} = 4.5$ obtained by formula (\ref{emt}). The difference between these values is probably due to the fact that in TM polarization the electric field has a component which is perpendicular to the $z$ axis.} $\varepsilon_\mathrm{eff, TE} \approx 3.4$ and $\varepsilon_\mathrm{eff, TM} \approx 4.2$.

To calculate the eigenmodes of the double slab waveguide we use the equation:
	\begin{equation}\label{eq11}
		\left(\frac{\xi_2}{\xi_1\xi_t}+\frac{\xi_1}{\xi_2\xi_i}\right)t_1t_2+q_1t_1+q_2t_2=\frac{1}{\xi_i}+\frac{1}{\xi_t}
	\end{equation}
where
\begin{equation}\label{eq111}
q_\alpha=\frac{1}{\xi_\alpha}-\frac{\xi_\alpha}{\xi_i\xi_t},
\end{equation}
$t_{\alpha}=\tan{k_{z\alpha}h_\alpha}$ ($\alpha=1,2$), $\xi_n=k_{zn}/\varepsilon_n$ for TM polarization and $\xi_n=1/k_{zn}$ for TE polarization; $k_{zn}$ stands for the $z$-component of wavevector in $n$-th medium, symbol $n$ denotes $i$, $1$, $2$ or $t$ which mean incoming medium, photonic crystal layer, non-modulated layer and outgoing medium correspondingly. The $z$-component of the wavevector $k_{zn}$ can be found from the equation $k_x^2+k_y^2+k_{zn}^2=\varepsilon_{n}k_0^2$, where $k_0$ is the absolute value of the photon wavevector in vacuum.

\begin{figure*}[t!]
\centering
\includegraphics[width=1\textwidth]{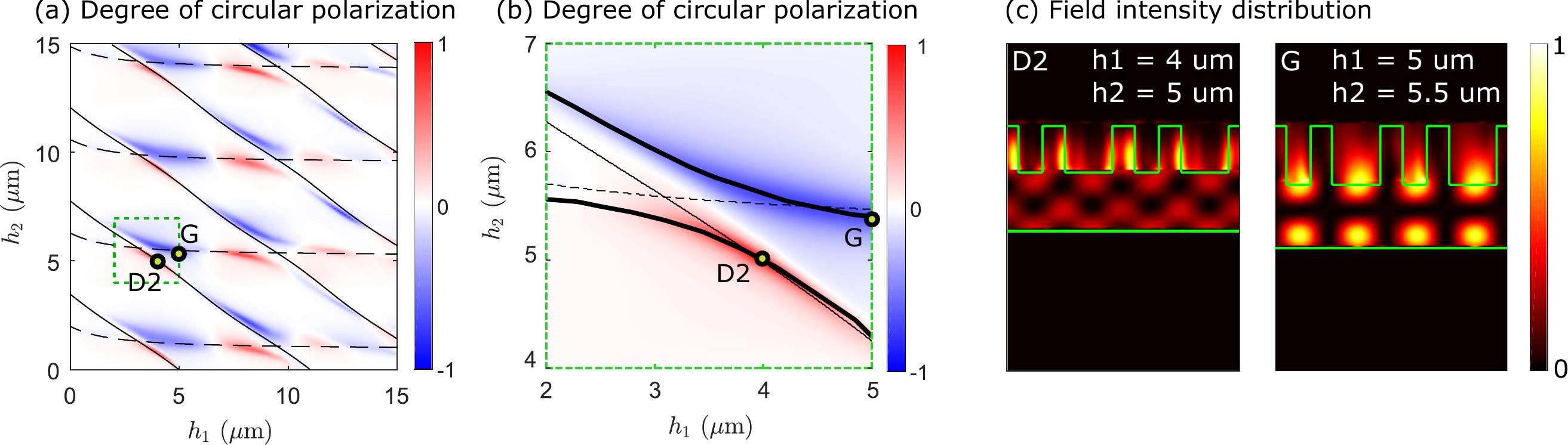}
\caption{(Color online) (a) Degree of circular polarization as a function of thicknesses $h_1$ and $h_2$. Resonances in empty lattice approximations are shown by dashed lines for TE polariztion and thin solid lines for TM polarization. The modes bounded by the green dashed rectangle are shown in the panel (b). Black thick lines in panel (b) denote resonances calculated for the initial chiral metasurface. (c): Electric field intensity of the chiral emitter eigenmodes calculated in the resonances D$_2$ and G specified in the panel (a).}
\label{fig3}
\end{figure*}

The dispersion of eigenmodes of the effective double slab waveguide (shown in Fig.\ref{fig2}b by black thick lines) are below the vacuum and substrate light cones and therefore are not visible from the far field. It is noteworthy that due to the different effective dielectric permittivities of UHM in TM and TE polarizations, $\varepsilon_\mathrm{eff,TM}$ and $\varepsilon_\mathrm{eff,TE}$, the TM guided mode appears to be below the TE guided mode, contrary to the case of isotropic waveguide. The introduction of periodicity folds the dispersion curves into the first Brillouin zone and couples the guided modes with photon continua in vacuum and substrate. In the result, so-called quasiguided modes appear \cite{Tikhodeev2002b}. It can be seen from Fig.\ref{fig2}b that several families of quasiguided modes are formed in $\Gamma$-point. One of these families is shown in Fig.\,\ref{fig2}c on a larger scale near $a/\lambda=0.78$.

Let us return to the initial periodic structure and demonstrate its eigenmodes and field distributions in them. To do this, we calculate the scattering matrix of the periodic structure, $\mathbb{S}$, which couples the incoming and outgoing amplitude vectors, $\ket{\mathbf{I}}$ and $\ket{\mathbf{O}}$, as defined in Ref.\,\onlinecite{Tikhodeev2002b}: 
\begin{equation}
\ket{\mathbf{O}}=\mathbb{S}(\lambda,k_x,k_y)\ket{\mathbf{I}}.
\end{equation}
By setting the incoming amplitudes as zero, we obtain the following eigenvalue problem \cite{Gippius2005c}:
\begin{equation}
\mathbb{S}^{-1}(\lambda,k_x,k_y)\ket{\mathbf{O}}_{res}=\ket{0}
\end{equation}
where $\ket{\mathbf{O}}_{res}$ is the resonant output eigenvector. The eigenmodes of the chiral metasurface are shown in Fig.\,\ref{disp} against the background of the DCP as a function of $k_x$ and $a/\lambda$. In comparison to the modes obtained in empty lattice approximation, the degeneracy of the modes in $\Gamma$ point is lifted except for two doublets B and D. Such modes picture is dictated by the structure symmetry C$_4$ \cite{Tikhodeev2002b}. The highest DCP is reached near doublets B and D. The electric field distributions in modes A--F are shown in Fig.\,\ref{efields}. The left panels in Fig.\,\ref{efields} demonstrate the field intensities. {On the right panels the electric field vectors are shown by the color cones featuring the field polarization characteristics. In such representation, the cone base denotes the polarization plane, while the cone height is proportional to the product of the electric field amplitude and the DCP. The color scale represents the phase of electromagnetic oscillations as explained in Fig.\,\ref{efields} (See supplemental materials for details).}

It is remarkable that in B and D doublets, where the structure emissivity is highly circularly polarized, the eigenfields are also circularly polarized.

As the next step, we calculate the DCP of thermal emission at $\lambda=13$\,$\mu$m as a function of the thicknesses of modulated and non-modulated parts of the structure, $h_1$ and $h_2$ (see Fig.\,\ref{fig3}a, red-blue image graph and the colormap on the right). It can be seen that the emissivity resonantly depends on the parameters $h_1$ and $h_2$. These resonances are attributed to the excitation of lossy quisiguided modes in the periodical ZnSe waveguide \cite{Tikhodeev2002b, lobanov2015polarization}. They can be approximately described in empty lattice approximation by taking the $x$ and $y$-components of wavevector in Eq.\,\ref{eq11} as $k_x=k_y=2\pi/a$ which  corresponds to the $\Gamma$-point in the higher order Brillouin zone of the reciprocal photonic crystal lattice. The solutions of transcendental Eq.\,\ref{eq11} are shown in Fig.\,\ref{fig3}a for TM polarization by dashed lines and for TE polarization by solid lines.

From Fig.\,\ref{fig3}a one can see that the resonances in TE and TM polarization have different behaviour with change of $h_1$ and $h_2$. Indeed, the TE quasiguided modes are practically not affected by the thickness of the photonic crystal slab $h_1$ while the TM modes depend both on $h_1$ and $h_2$. In terms of the empty lattice approximation, such distinction between the modes behaviour is originated from the UHM anisotropy. To explain this in more detail, we calculate the photon energies at $k_x=k_y=2\pi/a$ for the effective materials of effective double slab waveguide and compare these photon energies with $a/\lambda_0$ where $\lambda_0=13$\,$\mu$m (see Table\,\ref{table1}). By doing so, we fix the energy of guided modes at the value of $a/\lambda_0$ like in Fig.\,\ref{fig3}a. Inspection of Table\,\ref{table1} reveals that due to the different $\varepsilon_{eff,TM}$ and $\varepsilon_{eff,TE}$, the guided mode of the effective double slab waveguide appears to be above the UHM light cone in TM polarization and below the UHM light cone in TE polarization. This means that in the effective double slab waveguide, the TE guided modes are confined in the homogeneous part of the structure and hence depend only on thickness $h_1$. In contrast, the TM guided modes are confined both in the photonic crystal and homogeneous parts of the waveguide and depends on $h_1$ and $h_2$. 
\begin{table}
\label{table1}
  \caption{Effective dielectric permittivity of UHM and characteristic energies $a/\lambda$ of the effective double slide waveguide at $k_x=k_y=2\pi/a$ compared to $a/\lambda_0$.}
  \begin{tabular}{|l|cc|}
    \hline
   & TE & TM \\
	\hline
	    \hline
	effective epsilon of UHM  			& 3.4 & 4.5\\
    \hline
        \hline
light line in vacuum 		& \multicolumn{2}{c|}{1.414}\\
    \hline
    light line in substrate		& \multicolumn{2}{c|}{0.980}\\
    \hline
a/$\lambda_0$ 			& \multicolumn{2}{c|}{0.770}\\
    \hline
    light line in UHM 		& 0.771 & 0.665\\
    \hline
light line in ZnSe 		& \multicolumn{2}{c|}{0.594}\\
    \hline 
  \end{tabular}
  \label{table1}
\end{table}
Fig.\,\ref{fig3}b shows the electric field intensity distributions in the eigenmodes of the chiral metasurface with $h_1$ and $h_2$ which are close TE and TM resonances calculated in the empty lattice approximation (D$_2$ and G points in Fig.\,\ref{fig3}a). It can be seen that, indeed, the electric field is localized in the entire waveguide for the point D$_2$ and in the modulated part of the waveguide for the point G. Fig.\,\ref{fig3}c shows the eigenmodes of initial periodic structure on the $h_1-h_2$ diagram for $k_x=k_y=0$ and $\lambda=13$\,$\mu$m. It can be seen from Fig.\,\ref{fig3}b that the eigenmodes of periodic structure are close to the DCP maxima and are split near the intersection between TE and TM eigenmodes of the effective homogeneous double slab waveguide.

Finally, the angular emission diagram of thermal emission is shown in Fig.\,\ref{fig3}c for $\lambda=12.99$\,$\mu$m for two circular polarizations. We notice that the directionality of circular polarization is highly pronounced.

\begin{figure}[t!]
\centering
\includegraphics[width=0.6\columnwidth]{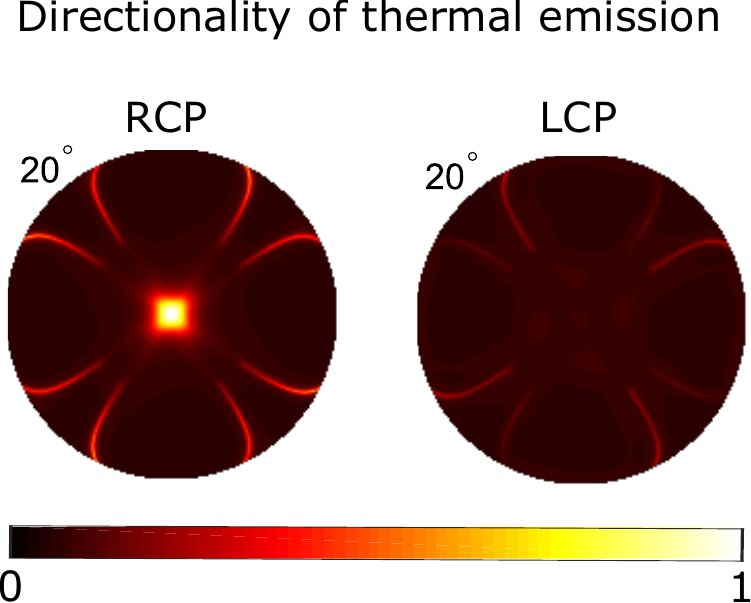}
\caption{(Color online) Directionality of thermal emission in right and left circular polarizations calculated for $h_1=4$\,$\mu$m, $h_2=5$\,$\mu$m.}
\label{fig4}
\end{figure}

\section{Conclusion}

In conclusion, we have demonstrated that the chiral metasurface can emit the circularly polarized thermal radiation with the degree of circular polarization as high as 87\%. We attribute this effect to the shape anisotropy of the photonic crystal part of the structure. The circularly polarized thermal emission is highly directional. 
 
This work has been funded by Russian Scientific Foundation (Grant No. 16-12-10538).

%

\end{document}